 \documentclass[aps,prl,twocolumn,groupedaddress,showpacs]{revtex4}
\usepackage{graphicx}       
\usepackage{dcolumn}        
\usepackage{bm}             

\begin{document}

\title
    {
        Direct observation of optically induced transient structures in graphite
        using ultrafast electron crystallography
    }

\author{Ramani K. Raman}
\affiliation{Physics and Astronomy Department,
             Michigan State University,
             East Lansing, Michigan 48824-2320 }

\author{Yoshie Murooka}
\affiliation{Physics and Astronomy Department,
             Michigan State University,
             East Lansing, Michigan 48824-2320 }

\author{Chong-Yu Ruan}
\email[]{Email: ruan@pa.msu.edu}
\affiliation{Physics and Astronomy Department,
             Michigan State University,
             East Lansing, Michigan 48824-2320 }

\author{Teng Yang}
\affiliation{Physics and Astronomy Department,
             Michigan State University,
             East Lansing, Michigan 48824-2320 }

\author{Savas Berber}
\affiliation{Physics and Astronomy Department,
             Michigan State University,
             East Lansing, Michigan 48824-2320 }

\author{David Tom\'anek}
\email[]{Email: tomanek@pa.msu.edu}
\affiliation{Physics and Astronomy Department,
             Michigan State University,
             East Lansing, Michigan 48824-2320 }



\begin{abstract}
We use ultrafast electron crystallography to study structural
changes induced in graphite by a femtosecond laser pulse. At
moderate fluences of ${\leq}21$~mJ/cm$^2$, lattice vibrations are
observed to thermalize on a time scale of ${\approx}8$~ps. At
higher fluences approaching the damage threshold, lattice
vibration amplitudes saturate. Following a marked initial
contraction, graphite is driven nonthermally into a transient
state with $sp^3$-like character, forming interlayer bonds. Using
{\em ab initio} density functional calculations, we trace the
governing mechanism back to electronic structure changes following
the photo-excitation.
\end{abstract}




\maketitle



There is growing interest in displacing atoms in materials by
photo-excitations \cite{Nasu2004}. Observations of transient
structures thus formed offer a glimpse into the transformation
pathways between different structures. Carbon, with its propensity
to form a wide range of bonding networks ($sp$, $sp^2$, $sp^3$) is
ideal to study the dynamics of bond formation and rupture. Of
particular interest is the conversion of graphite to diamond
\cite{MeguroAPL2001, NakayamaJPhysCM2003}, which is
believed to involve the rhombohedral phase of graphite as
intermediate state \cite{FahyPRB1986, GWYangApplPhysA2001}.
Whereas pioneering ultrafast optical studies of graphite have
provided evidence for photo-induced melting
\cite{AshitkovJETP2002, ReitzePRB1992} and generation of coherent
phonons \cite{MishinaPRB2000, IshiokaAPL2001} by observing changes
in the electronic properties, direct determination of lattice
structural dynamics from optical data proves difficult, especially
in the far-from-equilibrium regime.  X-ray diffraction has been
successful in observing nonthermal structural changes in
semiconductors \cite{BargheerChemPhysChem2006}, relying primarily
on the integrated Bragg intensities. However, direct observation
of atomic motion in nanostructures with low atomic number, such as
carbon, has not yet been achieved. Electron diffraction, with its
five orders of magnitude enhanced scattering cross-section and
advanced schemes achieving femtosecond temporal resolution
\cite{Zewail4DUEDReview2006, MillerFEDReview2006,
ParkSolidStCommun2005, RuanNanoLett2007}, offers a new window into
the realm of photo-excited structural dynamics, with sensitivity
down to ${\alt}1$~nm \cite{Zewail4DUEDReview2006,
RuanNanoLett2007, JanzenSurfSci2006}.

Here, we report the first direct determination of structural
changes induced in graphite by a femtosecond laser pulse. At
moderate fluences of ${\leq}21$~mJ/cm$^2$, after a short
thermalization period of ${\approx}8$~ps, we can attribute a
temperature value to the graphitic layers and find the interlayer
vibration amplitudes to match those reported in X-ray and neutron
scattering studies \cite{BTKelly1981}. At higher fluence values
approaching the damage threshold, we observe lattice vibration
amplitudes to saturate. Following a marked initial contraction of
the interlayer spacing by ${\leq}6\%$, graphite is driven
nonthermally into a transient state with $sp^3$-like character,
forming $1.9$~{\AA} long interlayer bonds. Using {\em ab initio}
density functional theory (DFT) calculations, we trace the
structural changes back to a nonthermal heating of the electron
gas, followed by a photo-induced charge separation causing a
compressive Coulomb stress.

\begin{figure}
\includegraphics[width=1.0\columnwidth]{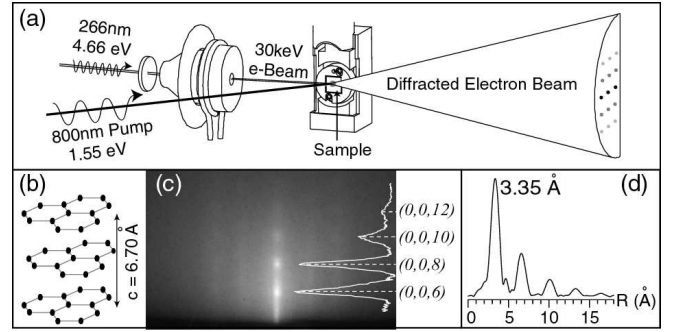}
\caption{ Ultrafast Electron Crystallography (UEC) of highly
oriented pyrolytic graphite. (a) The UEC pump-probe setup. (b) The
layered structure of graphite. (c) Ground state diffraction
pattern of graphite. (d) Ground state layer density distribution
function (LDF), obtained via a Fourier transform of the central
streak pattern in (c).\label{Fig1}}
\end{figure}

The experimental setup, shown in Fig.~\ref{Fig1}(a), has been
described in detail elsewhere \cite{RuanNanoLett2007}. Freshly
cleaved samples of highly oriented pyrolytic graphite (HOPG) were
placed inside an ultrahigh vacuum chamber through a load lock
system at room temperature. A mode-locked Ti-Sapphire laser
($p-$polarized, $45$~fs, $800$~nm, $1$~kHz) generated the pump
pulse, which was marginally focused (${\approx}600$~$\mu$m
diameter) onto the sample at ${\approx}45^{\circ}$ relative to the
HOPG c-axis. The range of the laser fluence $0.5<F<90$~mJ/cm$^2$
was below the {\em in situ} determined optical damage threshold of
${\approx}120$~mJ/cm$^2$.

A photo-generated $30$~keV electron beam (wavelength
$\lambda_e=0.069$~{\AA}), with diameter demagnified to
${\leq}5$~$\mu$m, was directed onto the sample at grazing
incidence of $2.4^{\circ}$ to serve as the probe. A translational
optical delay stage was used to vary the relative delay between
the arrival of pump and probe pulses at the sample, to observe the
changes induced by laser irradiation. The electron pulse sampled
the top $3-5$ layers of the graphite structure, depicted in
Fig.~\ref{Fig1}(b), and produced oriented molecular diffraction
patterns in the central streak region, shown in
Fig.~\ref{Fig1}(c). The layer density distribution function (LDF),
shown in Fig.~\ref{Fig1}(d), was obtained from the Fourier
analysis of the interference pattern. The primary peak at
$3.35$~{\AA} and less pronounced peaks at $6.7$~{\AA} and
${\approx}10$~{\AA} were found to be in good agreement with the
layered structure of bulk graphite \cite{BTKelly1981}. The decay
of higher order LDF peaks suggest a probing depth of ${\approx}1$~nm.

\begin{figure}
\includegraphics[width=1.0\columnwidth]{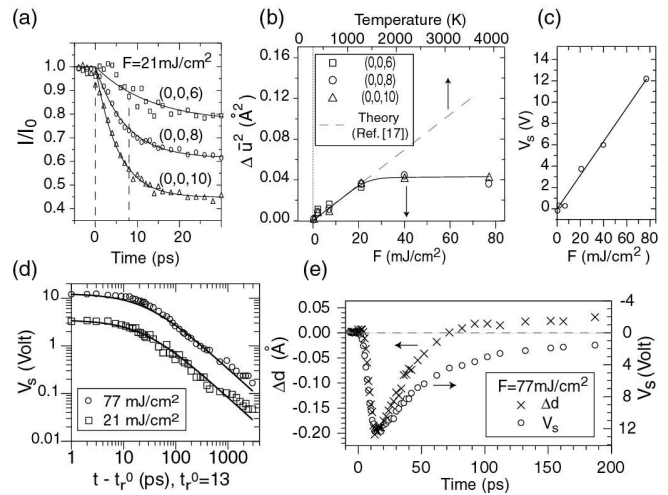}
\caption{ (a) Decay of the diffraction intensity. (b) Change in
mean-squared atomic displacements perpendicular to the graphite
planes, estimated from the Debye-Waller analysis of the intensity
drops in (a). (c) Effective surface potential $V_s$ at the HOPG
surface as a function of the applied fluence $F$. (d) Decay of the
effective surface potential with fit to the drift-diffusion
recombination model (see text). (e) Contraction
of the interlayer spacing ${\Delta}d$ (left axis) and the
corresponding increase in the effective surface potential (right
axis) in the far-from equilibrium regime.
\label{Fig2}}
\end{figure}

To study the photo-induced structure dynamics, we first examine
the near-equilibrium regime at low fluences of
0.5~mJ/cm$^2{\leq}F{\leq}21$~mJ/cm$^2$. Random atomic
displacements diminish the intensity of the diffraction spectra by
a Debye-Waller factor $e^{-2M}$, where $M=-s^2\bar{u}^2/4$,
$\bar{u}^2$ is the mean-square atomic displacement perpendicular
to the reflecting planes, and $s=(4\pi/\lambda_e)\sin(\theta/2)$
is the momentum transfer associated with the maxima located at the
scattering angle $\theta$. Thus, the change in the mean-square
atomic displacement ${\Delta}\bar{u}^2$, measured relative to the
unperturbed state at negative times, determines the rise in
lattice temperature and can be calculated from the diffraction
intensities as ln$\left(I_s(t)/I_s(t<0)\right)=-s^2\Delta
\bar{u}^2/4$. In Ultrafast Electron Crystallography (UEC), we
monitor the integrated intensity of the (0,0,6), (0,0,8) and the
(0,0,10) maxima arising from the interference between the
graphitic layers, shown in Fig.~\ref{Fig2}(a). The intensity of
all 3 maxima drops within a timescale of $8{\pm}1$~ps, indicative
of increased thermal motion of the graphitic planes. As we probe
reflections from the basal planes, the observed intensity drops
mainly due to the increased out-of-plane displacement of the
atoms. Recent optical studies \cite{MishinaPRB2000,
IshiokaAPL2001} have associated near-infrared optical excitations
in graphite with a femtosecond generation of coherent phonons with
$E_{2g}$ symmetry (interlayer shearing mode). Hence, this $8$~ps
timescale of interlayer thermal excitation is a direct measure of
the phonon-phonon interactions in HOPG, and is in good agreement
with reports of hot phonon relaxation times in graphite
\cite{KampfrathPRL2005}.

Quantitative analysis of the intensity drops in Fig.~\ref{Fig2}(a)
indicates that the mean square atomic displacement
$\Delta\bar{u}^2$ increases linearly with the applied fluence in
the near-equilibrium regime, as seen in Fig.~\ref{Fig2}(b).
Comparing our data to a theoretical model of temperature dependent
$\Delta\bar{u}^2$, benchmarked with X-Ray and neutron scattering
data \cite{BTKelly1981}, we estimate a temperature rise of 950~K
at $F=21$~mJ/cm$^2$ in Fig.~\ref{Fig2}(b). This is also in good
agreement with temperature extracted from optical studies using
heat capacity and absorbance \cite{ReitzePRB1992}.

As we enter the far-from-equilibrium regime by increasing the
excitation fluence beyond $21$~mJ/cm$^2$, the amplitude of atomic
vibrations saturates and no longer increases with increasing
fluence, as seen in Fig.~\ref{Fig2}(b). This is rather surprising,
for one would expect a monotonic increase in the thermal motion as
graphite approaches its melting threshold at $130$~mJ/cm$^2$
\cite{ReitzePRB1992}. Furthermore, the (0,0,6) maximum is found to
shift by almost 0.34 ~{\AA}$^{-1}$ at $F=40$~mJ/cm$^2$, which
would correspond to an interplanar expansion of 6\%. Such a large
peak shift cannot arise from structural changes alone.

We believe that the key effect of the optical excitation is the
repopulation of electronic states, causing re-bonding and
redistribution of charge at the surface\cite{MurdickPRB2008}. This
charge redistribution gives rise to a Coulomb field near the
graphite surface, which induces a collective shift of the electron
diffraction pattern. We measure this Coulomb field here directly
by modeling the `refraction' effect commonly associated with the
existence of a surface potential at the vacuum-material interface
\cite{MurdickPRB2008}. We deduce the effective surface potential
$V_s$ as a function of the applied fluence and find it to increase
linearly up to $F{\approx}80$~mJ/cm$^2$, as shown in
Fig.~\ref{Fig2}(c). This linear trend indicates that the surface
charging is mainly driven by the non-equilibrium diffusion of
photogenerated carriers rather than multiphoton ionization. The
decay of this transient surface potential resulting from the
recombination of space charge follows a power-law decay with the
exponent $-1$, as shown in Fig. 2(d) and predicted by the
drift-diffusion recombination model \cite{MurdickPRB2008}.

To examine the effect of optical excitations on the atomic
structure at these elevated fluences, we monitored the interlayer
separation following the photo-excitation by a time-resolved LDF
analysis, after having accounted for the surface potential
effects. In contrast to the thermal expansion at low fluences, we
observed the interlayer separation to contract rapidly, as seen in
Fig.~\ref{Fig2}(e). The change in the interlayer spacing $\Delta$d
is found to correlate well with the rise and fall of $V_s$ at
$F=77$~mJ/cm$^2$, with a maximum value $V_s\approx$~12 V
corresponding to an interlayer contraction of $\approx6\%$. The
potential rise $V_s$ sampled within the electron probe depth of 1
nm yields an internal field of $E\approx1.2$ V/~{\AA}, which
causes Coulomb stress. Using this value of $E$ and $\epsilon_r=10$
for graphite, we estimate a maximum energy density of
$U_e={\epsilon_r}{\epsilon_0}E^2/2=0.2$~eV/atom, close to the
${\approx}0.3$~eV/atom value based on DFT for the activation
barrier in the graphite-to-diamond transformation
\cite{FahyPRB1986, NakayamaJPhysCM2003}.

\begin{figure}
\includegraphics[width=1.0\columnwidth]{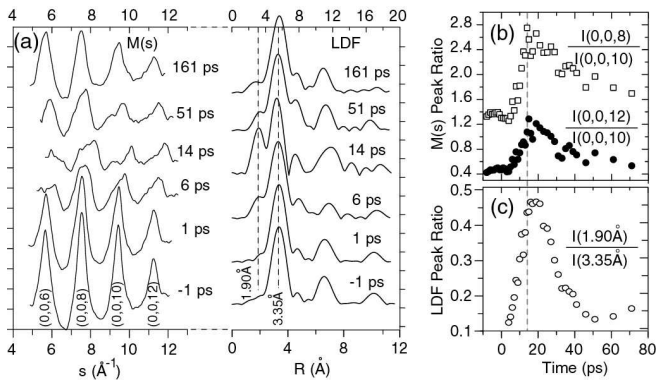}
\caption{ Signature of $sp^3$-like bonding in HOPG. (a) Molecular
interference pattern $M(s)$ and the corresponding LDF curves at
selected time stances following strong photo-excitation. The
transient peak at $1.9$~{\AA} in the LDF indicates formation of an
interlayer bond. (b) Ratio of peak intensities obtained from the
diffraction patterns. The enhancement of the $(0,0,12)$ with
respect to the $(0,0,10)$ peak cannot be explained by thermal
vibrations and indicates a transient structure consistent with
$sp^3$ bonding. (c) Ratio of peak intensities at $R=1.9$~{\AA} and
$3.35$~{\AA} in the LDF. \label{Fig3}}
\end{figure}

To further elucidate the structural evolution, we examined closely
the diffraction spectra. Figure~\ref{Fig3}(a) shows the time
evolution of the molecular interference curves $M(s)$ and the
corresponding LDF curves at $F=77$~mJ/cm$^2$, just below the
damage threshold. Starting at ${\approx}6$~ps, as $V_s$ rises
beyond 4~V, we observe the onset of peak shouldering and
broadening in the $M(s)$ curve and appearance of a new peak at
$R{\approx}1.9$~{\AA} in the LDF curve. Concurrent with this, we
also observe a drastic change of relative intensities between the
even and odd peaks in the $M(s)$ curves. In graphite with the
interlayer spacing of $d_0=c_0/2=3.35$~{\AA}, the even Bragg peaks
$n=4,6$ are labeled $(0,0,8)$ and $(0,0,12)$, and the odd peaks
$n=3,5$ are labeled $(0,0,6)$ and $(0,0,10)$, respectively. The
observed higher intensity of the $(0,0,12)$ peak with respect to
the $(0,0,10)$ peak in Fig. \ref{Fig3}(b) cannot be explained by
thermal vibrations and suggests a significant structural change.
Consistent with the fact that only the even diffraction peaks
should be present in the structure of diamond, and also with the
observation of a well-defined interlayer bond at
$R{\approx}1.9$~{\AA}, we conclude that the transient structure is
consistent with $sp^3$ bonding. We find the transient $sp^3$-like
structure, associated with the relative intensity of the
$R{\approx}1.9$~{\AA} peak in the LDF curve, to reach a maximum
fraction of 47\% at 14~ps, as seen in Fig. \ref{Fig3}~(c). Similar
$sp^2-sp^3$ hybrid structures have also been identified in
pressure-driven pathways of graphite-to-diamond
transition\cite{MaoScience2003}. By 45~ps, the structure recovers
its $sp^2$ character, but remains hot even after $3$~ns with an
average value ${\Delta}\bar{u}^2=0.033$~{\AA}$^2$, corresponding
to a lattice temperature $T{\approx}1000$~K.

\begin{figure}
\includegraphics[width=0.95\columnwidth]{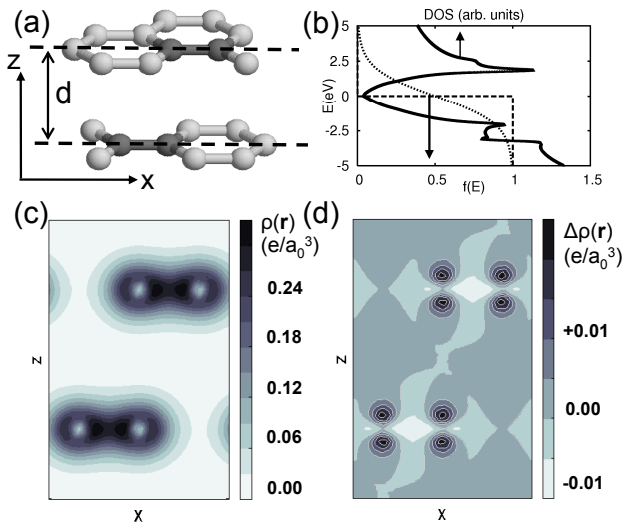}
\caption{(Color online)
Electronic structure changes during photo-excitations in graphite.
(a) Schematic view of the structure.
(b) Electronic density of states (solid line) and the Fermi-Dirac
distribution of graphite at $T=0$ (dashed line) and
$k_BT=1.0$~eV (dotted line).
(c) Total pseudo-charge density $\rho({\bf r})$ and
(d) change in the pseudo-charge density $\Delta\rho({\bf r})$
due to an electronic temperature increase to $k_BT=1.0$~eV.
The plane used in (c) and (d) intersects the graphene layers
along the dashed lines in (a). \label{Fig4}}
\end{figure}

To clarify the origin of the structural changes, we studied the
effect of laser pulses on the electronic structure and bonding in
hexagonal graphite using {\em ab initio} DFT calculations in the
local density approximation (LDA). While much of the basic physics
underlying photo-induced structural changes can be understood in
the bulk structure, we used graphite slabs representing the
surface for quantitative predictions. The total energy was
determined using the ABINIT plane-wave code \cite{ABINIT} with a
64~Ry energy cutoff, Troullier-Martins pseudopotentials, and the
Ceperley-Alder form of the exchange-correlation functional. The
Brillouin zone of the $4-$atom bulk unit cell was sampled using a
fine mesh of $24{\times}24{\times}12$ $k$-points that included the
$K-H$ line at the Brillouin zone edge, close to the Fermi surface.
This approach correctly reproduced the observed \cite{BTKelly1981}
in-layer bond length $d_{CC}=1.42$~{\AA} and the interlayer
spacing $d_0=3.34$~{\AA} in the bulk system at $T=0$, shown
schematically in Fig.~\ref{Fig4}(a). The electronic density of
states of graphite near the Fermi level is shown in
Fig.~\ref{Fig4}(b), and the total charge density is depicted in
Fig.~\ref{Fig4}(c) in a plane normal to the layers. In the
following, we will study the effect of two types of photo-induced
electronic excitations on the structure.

First, we will consider the thermalization of the initial
nonequilibrium population of electronic states in the
laser-irradiated target over a sub-picosecond time scale
\cite{MiyamotoPRL2006} to that of a very hot electron gas
\cite{SilverstrelliPRL1996}. The electronic temperature ($T_e$) is
probably lower than the limiting value $k_BT_e{\alt}h\nu=1.55$~eV
and depends on the laser fluence. For the sake of illustration, we
compare the Fermi-Dirac distribution in graphite at $k_BT_e=0$ and
$k_BT_e=1.0$~eV in Fig.~\ref{Fig4}(b), indicating the electronic
excitations in the $\pi-\pi^*$ manifold of the hot electron gas.
Populating initially empty conduction states by valence electrons
leads to a change in the total electron density
${\Delta}{\rho}({\bf r})={\rho}({\bf r};k_BT_e)-{\rho}({\bf
r};0)$, which is depicted in Fig.~\ref{Fig4}(d) for
$k_BT_e=1.0$~eV.

As stipulated by DFT, changes in the charge density modify the
force field in the system. Our results in Fig.~\ref{Fig4}(d) suggest
an increased population of C$2p_z$ orbitals that may hybridize to
$pp\sigma$ bonding states connecting neighboring layers, increasing their
attraction. The depopulation of in-layer bonding states, on the other hand,
should cause an in-layer expansion. To obtain a quantitative estimate of
photo-induced structural changes, we performed a set of global structure
optimization calculations of graphite with
electrons subject to effective temperatures in the range
$0$~eV$<k_BT<1.55$~eV. We used a stringent convergence criterion,
requiring that all stress components lie below
$5.0{\times}10^{-7}$~Ha/a$_0^3$ and that no force exceeds
$5.0{\times}10^{-5}$~Ha/a$_0$. In bulk graphite, we indeed found
that nonzero electronic temperature leads to a maximum interlayer
contraction ${\Delta}d/d_0=-1$\% and in-layer expansion
${\Delta}r_{CC}/r_{CC}=+1$\% for $k_BT_e{\approx}1.0$~eV.
In the slab geometry, we obtained a slightly larger interlayer
contraction of up to ${\Delta}d/d_0=-1.5$\%.

To further take into account the effect of Coulomb stress induced
by the laser pulse, we represent the charge separation in the
electronic  ground state, corresponding to the observed internal
field of ${\approx}1.2$~V/{\AA}. To model this system, we immersed
three-layer graphite slabs, separated by $30$~{\AA}, in a uniform
electric field and adjusted its strength to reproduced the
observed field value. With the polarized charge distribution
frozen in, the external electric field was switched off. Then, the
internal field was only due to the polarized charge distribution.
The slab calculations were performed using the SIESTA code
\cite{SolerJPhysCM2002} with double-polarized triple-$\zeta$ local
basis and a fine $24{\times}24$ $k$-point mesh. The charge density
was obtained on a real-space grid with a mesh cutoff energy of
250~Ry. From total energy calculations at different interlayer
separations, we found that charge density redistribution
associated with the internal field may further reduce the
interlayer separation by $2-3$\%. Combined with the contraction
induced by the initial non-equilibrium electron heating, we thus
can explain the observed contraction of the topmost interlayer
separation by up to ${\approx}5$\%.

In conclusion, a nonthermal pathway of photo-induced structural
changes in graphite has been observed through multidimensional
crystallographic determination in space, time and energy. Beyond a
threshold fluence, a transient $sp^3$-like structure with
well-defined interlayer bonds emerges. The modified force field in
the excited state and the Coulomb stress are the main forces
driving this structural change.

This work was supported by Department of Energy under grant
DE-FG02-06ER46309 (RKR, YM, and CYR), and by the National Science
Foundation under NSF-NSEC grant 425826 and NSF-NIRT grant
ECS-0506309 (SB, TY and DT).


\end{document}